\documentclass[prd,twocolumn,amsmath,aps,floats,amssymb, floatfix, superscriptaddress, nofootinbib]{revtex4}

\pdfoutput=1

\usepackage{graphicx}
\usepackage{wasysym}
\usepackage{amsfonts}
\usepackage{subfigure}
\usepackage{hyperref}
\usepackage{color}        
\usepackage{ulem}
\usepackage[usenames,dvipsnames]{xcolor}

\begin{document}

\title{Reionization history and CMB parameter estimation}

\author{Azadeh Moradinezhad Dizgah}
\affiliation{Department of Physics, University at Buffalo, The State University of New York, Buffalo, NY 14260-1500, USA}

\author{Nickolay Y.\ Gnedin}
\affiliation{Particle Astrophysics Center, Fermi National Accelerator Laboratory,
Batavia, IL 60510, USA}
\affiliation{Department of Astronomy \& Astrophysics, The University of
Chicago, Chicago, IL 60637 USA}
\affiliation{Kavli Institute for Cosmological Physics and Enrico Fermi Institute,
The University of Chicago, Chicago, IL 60637 USA}

\author{William H.\ Kinney}
\affiliation{Department of Physics, University at Buffalo, The State University of New York, Buffalo, NY 14260-1500, USA}

\begin{abstract}
We study how uncertainty in the reionization history of the universe affects estimates of other cosmological parameters from the Cosmic Microwave Background. We analyze WMAP7 data and synthetic Planck-quality data generated using a realistic scenario for the reionization history of the universe obtained from high-resolution numerical simulation. We perform parameter estimation using a simple sudden reionization approximation, and using the Principal Component  Analysis (PCA) technique proposed by Mortonson and Hu. We reach two main conclusions: (1) Adopting a simple sudden reionization model does not introduce measurable bias into values for other parameters, indicating that detailed modeling of reionization is not necessary for the purpose of parameter estimation from future CMB data sets such as Planck. (2) PCA analysis does not allow accurate reconstruction of the actual reionization history of the universe in a realistic case.
\end{abstract}

\maketitle

\section{Introduction}
Temperature and polarization power spectra of cosmic microwave background (CMB) currently offer the most precise test of cosmological parameters. However the CMB is subject to cosmological foregrounds. If there is a phase of the history of the universe that we do not understand completely, it may bias our cosmological interpretation of the CMB. One such a phase is the epoch of reionization (EoR). Given the enhanced precision of upcoming Planck data, understanding and quantifying this bias is of significant importance. List of previous studies on this topic includes but not limited to \cite{Zahn:2005fn, Mortonson:2007tb, Mortonson:2008rx, Colombo:2008jr,Pandolfi:2010dz, Pandolfi:2010mv}. 

Reionization of neutral hydrogen is the last major thermodynamic phase transition in the history of the universe. This era begins by the emergence of the first luminous sources such as early star-forming galaxies and rare quasars and it is believed to be completed
a hundred million years later. As the luminous sources start to produce energetic photons, they reionize the neutral hydrogen around them. Free electrons generated during reionization will Thompson scatter the CMB photons in the period after recombination. This scattering results in small suppression in CMB temperature anisotropy on all scales and also will generate an extra polarization signal on large angular scales. A commonly used parameter to characterize these two effects is the total scattering optical depth $\tau$. The current constraint on the value of $\tau$ from WMAP7 \cite{Larson:2010gs} derived from the measurement of the amplitude of E-mode of polarization is $\tau = 0.087 \pm 0.017$. 

The above constraint on $\tau$ is obtained assuming that reionization is a sharp transition corresponding to redshift of $z_{reion} = 10.5 \pm 1.2$. However a sudden reionization requires an infinite photon production rate. Since the galaxy mass function evolves smoothly, reionization should be a gradual process which takes of order of Hubble time. Therefore this unphysical assumption about the reionization history can introduce a bias in the constraints on $\tau$ and other cosmological parameters degenerate with it. Another possible source of bias arises if temperature and polarization spectra of CMB are sensitive to details of reionization history beyond the optical depth. In this case, even if modeling reionization as a sudden process gives the correct estimate of $\tau$, inaccurate modeling of redshift evolution of number density of free electrons in intergalactic medium, $x_e$, would result in an additional bias.

In recent years there have been a significant improvements in numerical simulations and semi-analytical models of reionization which provide us with a more realistic picture of this era. In the first part of our analysis we adapt the result of one of these simulations to investigate the interplay between reionization history and CMB parameter estimation. Although the combination of simulation and semi-analytical models provide us with a more physical picture of EoR, the reionization process is complex, and there is currently a lack of direct observational data on the EoR. Therefore, an accurate model-independent approach to reconstruction of the reionization history is also of great importance.  The second part of our analysis is focused on the study  of accuracy of the commonly used model-independent approach (Hu $\&$ Holder, Mortonson $\&$ Hu) for Planck-level sensitivity data. 

The  outline of the paper is as following: in Section \ref{sec:reionmodel} we discuss the more realistic reionization model that we have adapted in our analysis, in Section \ref{sec:analysis} we discuss the analysis methods while in Section \ref{sec:results} we present our results. In Section \ref{sec:conclusions} we draw our conclusions. 

\section{Constraints on reionization history from simulations}
\label{sec:reionmodel}
\begin{figure}
\includegraphics[width=0.45\textwidth,clip]{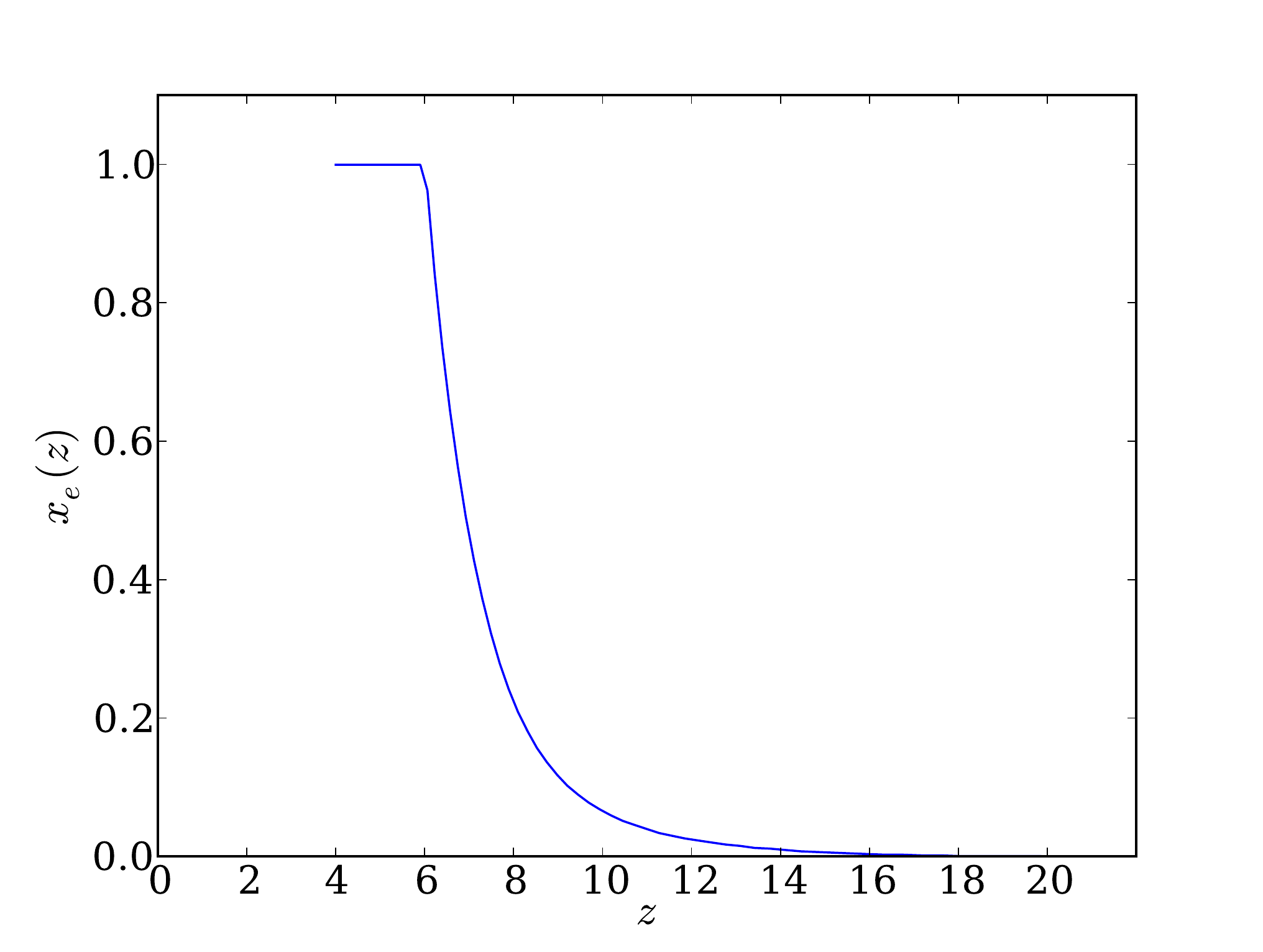}
\caption{Reionization history from Volenteri $\&$ Gnedin \cite{Volonteri:2009ck} }
\label{fig:Nick-reion}
\end{figure}

We adapt the reionization history from Volonteri $\&$ Gnedin \cite{Volonteri:2009ck} as a physically motivated reionization model. The ionized fraction as a function of redshift for this model is shown in Fig. \ref{fig:Nick-reion}. This result is based on theoretical modeling of sources of ionizing photons and counting the number of
ionizing photons per atom, while accounting the photon loss due to recombination in an approximate manner. The photon budget in this model is supplied by two main sources: stellar population and quasars. When considering the contribution of quasars to the total reionizing photons, the ``secondary reionization'' by energetic 
photons from quasars is also included.  Other more exotic sources such as energetic photons and electrons from dark matter annihilation (see \cite{Belikov:2009qx} and references therein) are neglected. 

The contribution from quasars to the total ionizing photons is estimated by modeling the evolution of massive black holes within dark matter halos.  The contribution from the star-forming galaxies to the total ionizing photons is calculated by extrapolating of the observed UV luminosity functions of high redshift galaxies (Bowens, {\it{et al.}}\cite{Bouwens:2007hn}, \cite{Bouwens:2008hm}) and using an estimate of relative escape fraction of the ionizing radiation from Gnedin (2007) \cite{Gnedin:2007pw}. We refer the reader to Volenteri $\&$ Gnedin \cite{Volonteri:2009ck} for the details of this calculation; here we only list the main ingredients of the model for the sake of completeness.

The total number of ionizing photons per atom is given by:
\begin{equation}
N_{\gamma/a} = N_{\gamma/{a,*}} + \frac{U_\gamma}{E_\gamma n_a} + f_{SI}(x) \frac{U_\gamma}{14.4eV n_a}.
\end{equation}
The first term in the above equation accounts for the contributions from stars. The second term is the contribution from ionizing photons from quasars with the mean energy of $E_\gamma$ while the last term includes the secondary ionizations from energetic photons. Quantity $n_a = (1-0.75Y_p)n_b \approx 2.0 \times 10^{-7} cm^{-3}$ is the comoving number density of hydrogen or helium atoms, 14.4 eV is the mean ionization energy per atom, and $U_\gamma$ is the total comoving energy density of ionizing radiation emitted by the population of quasars. The mean photon energy $E_\gamma$ is a parameter of the model and is assumed to be $300 \textrm{eV}$. The exact value of $E_\gamma$ depends on the specific shape of the average quasar spectrum. However, since most of the contribution from quasars comes from secondary ionizations, which are independent of $E_\gamma$, the result is only mildly affected by the actual value of that parameter (see \cite{Volonteri:2009ck} for more detailed discussion).  For the uniformly ionized gas, $f_{\textrm{SI}}$ is the fraction of  radiation energy density going into secondary ionizations, which is calculated in \cite{1985ApJ...298..268S} and fitted by a simple formula in \cite{Ricotti:2001zg},
\begin{equation}
f_{SI}(x) \approx 0.35(1-x^{0.4})^{1.8}-1.77\left(\frac{28eV}{E_\gamma}\right)^{0.4} x^{0.2}(1-x^{0.4})^2.
\end{equation}
Uniform ionization is a simplistic assumption. Therefore the quantity $x$ is an effective value $x_{\textrm{eff}}$ such that
\begin{equation}
N_{\gamma/a}(x_{\textrm{eff}}) = \left<N_{\gamma/a}(x)\right>
\end{equation}
and the average is mass-weighted. A complete calculation of $x_{\textrm{eff}}$ requires precise modeling of transfer of radiation throughout the universe. In the absence of such modeling, Volonteri, {\it{et al.}} assumed the following ansatz:
\begin{equation}
x_{\textrm{eff}} = \textrm{min}(1,f_{n \rightarrow x}N_{\gamma/a}),
\end{equation}
where $f_{n \rightarrow x}$ is a free parameter that accounts for loss of ionizing photons due to recombinations of previously ionized atoms. In reality it may be a function of time; it is unlikely that $f_{n \rightarrow x}$ is much smaller than one, or the universe would not be fully ionized by $z \approx 6$. For the reionization history shown in Fig. \ref{fig:Nick-reion}, $f_{n \rightarrow x} = 0.75$ (our fiducial value). 

\section{Analysis} 
\label{sec:analysis}
We investigate the dependence of constraints on cosmological parameters obtained from CMB on the assumed reionization history. We consider three reionization histories: instantaneous reionization, a physically motivated reionization history described in Sec. \ref{sec:reionmodel} (Hereafter we refer to this reionization history as VG model) and a general reionization history parameterized in terms of its principal components (PCs) with respect to E-mode polarization (Hu $\&$ Holder, Mortonson $\&$ Hu). We perform three sets of Markov Chain Monte Carlo (MCMC) analyses using the publicly available CosmoMC code \cite{Lewis:2002ah} as modified by Mortonson $\&$ Hu \cite{Mortonson:2007hq} and compare the constraints on cosmological parameters obtained for the above reionization profiles. 
\begin{figure*}
\centering
$\begin{array}{ccc}
\subfigure[]{
\includegraphics[width=0.52\textwidth,clip]{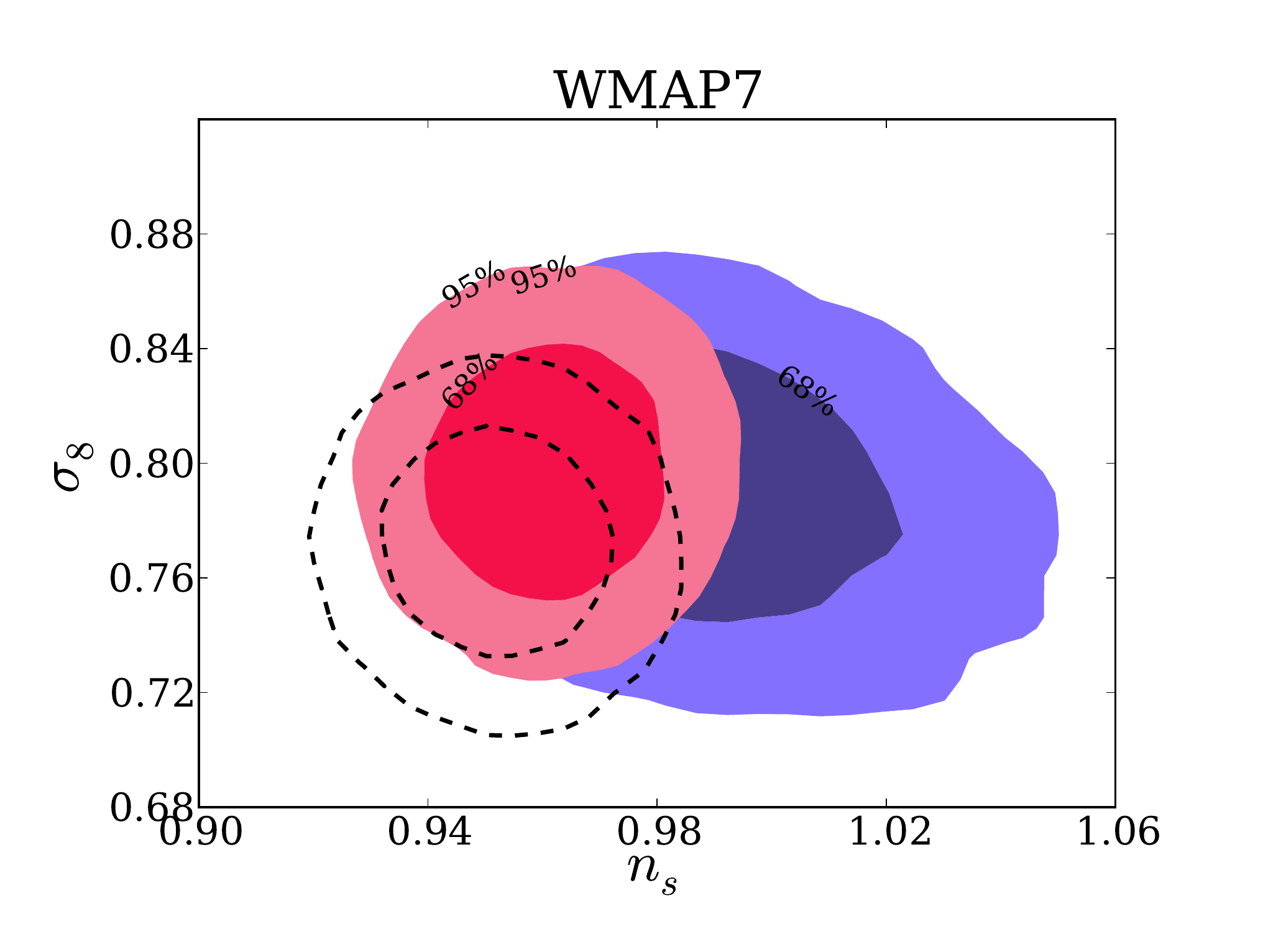}}
\subfigure[]{
\includegraphics[width=0.5 \textwidth,height=7cm,clip]{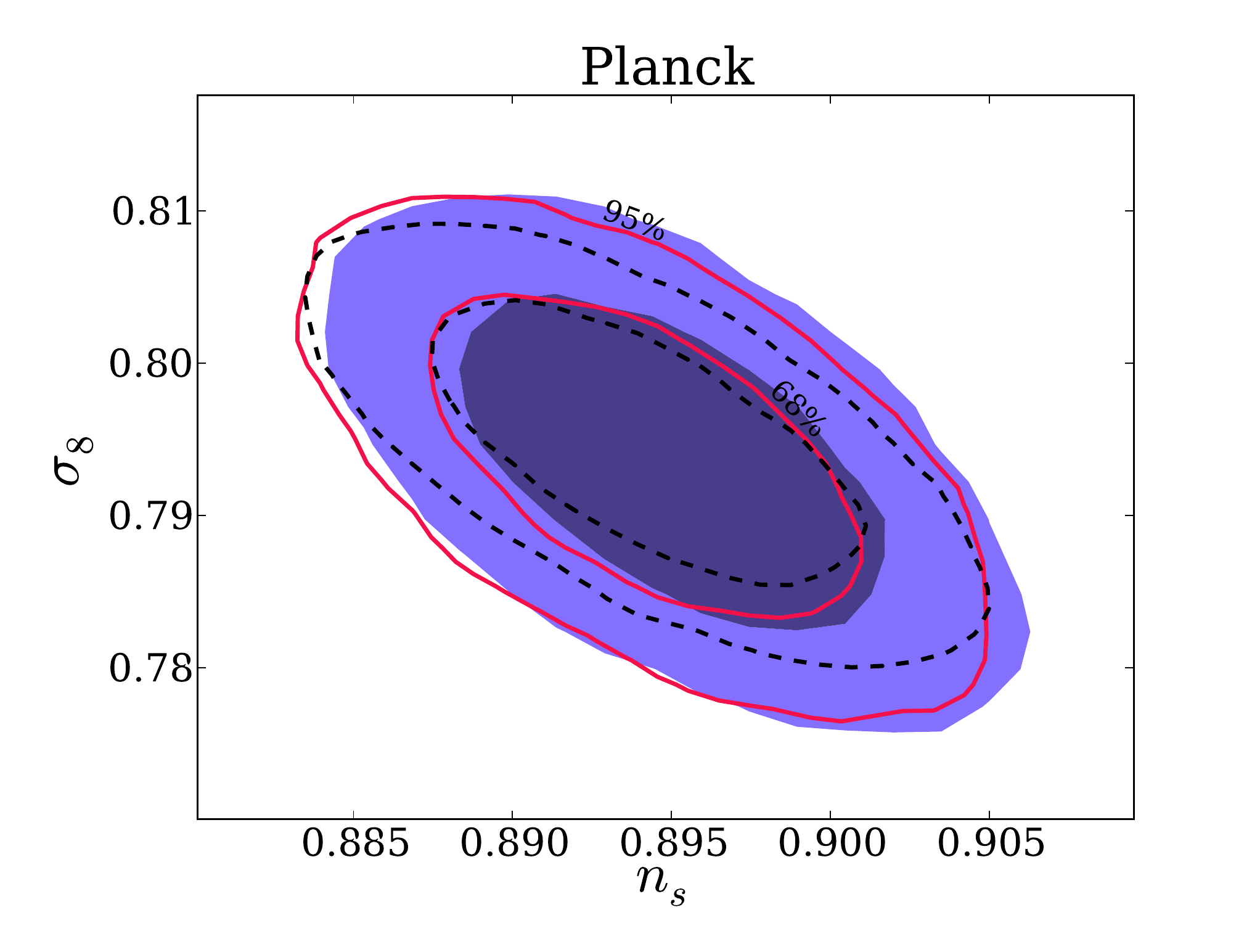}}
\end{array}$
\caption{Marginalized constraints in the $\sigma_8$-$n_s$ plane for (a) WMAP7 data and (b) simulated Planck data sets. Shaded blue contours correspond to PC reionization while dotted contours are obtained assuming VG reionization history. In (a) the shaded red contours correspond to sudden reionization while in (b) the constraints for sudden reionization are shown with solid red contours.}
\label{fig:sigma8-ns}
\end{figure*}

To consider a general reionization history, the ionizaed fraction, $x_e$ is considered as a free function of redshift and is decomposed into its principal components (PCs) with respect to E-mode polarization of the CMB:
\begin{equation}
x_e(z) = x_e^{fid}(z) + \sum \limits_i m_i S_i(z).
\vspace{-5pt}
\label{eq:PCA}
\end{equation}
$S_i$ are the eigenvalues of the fisher matrix that describes the dependance of $C_l^{EE}$ on $x_e(z)$ and $m_i$ are the amplitudes of the principal components. $x_e^{fid}$ is the fiducial model at which the Fisher matrix is calculated.  The principal components are defined over a limited range in redshift, $6<z<30$, with $x_e =0$ at redshifts $z>30$ and $x_e = 1$ at redshifts $z<6$. In our MCMC analysis we use the five lowest-variance principal components constructed around a constant reionization history of $x_e^{fid} = 0.15$ and vary the amplitudes of these components. We impose the same priors on $m$ as those in Mortonson $\&$ Hu \cite{Mortonson:2007hq}.

The data sets used are temperature and polarization data from 7-year WMAP (WMAP7) \cite{Komatsu:2010fb} and simulated Planck-precision data described below. We consider a flat $\Lambda$CDM cosmology with the base cosmological parameters chosen to be the baryon and cold dark matter densities, $\Omega_b h^2$ and $\Omega_c h^2$,  the ratio of the sound horizon to the angular diameter distance at decoupling, $\theta_s$ and the spectral index of scalar perturbations, $n_s$.  Following Mortonson and Hu, instead of varying $A_s$, we take a combination of $10^ {10} A_s e^{-2\tau}$ as a free parameter. This removes the known degeneracy between the amplitude of scalar perturbations and the optical depth. The optical depth, $\tau$, is then not a free parameter but is a derived parameter. For the general reionization parameterized by principal components, the five lowest variance amplitudes of PCs are varied. For WMAP7 data, the priors on cosmological parameters are set to the best fit values from the WMAP team. To have self-consistent comparison, for Planck data, the priors on the base cosmological parameters are set to best-fit values of our WMAP7 run with VG reionization history.  Chain convergence was determined with the Gellman-Rubin statistic, $R-1<0.1$. 

Planck data is simulated for a fiducial model with the best fit values from our WMAP7 run with VG reionization history with an optical depth of $\tau = 0.0408$.  Since for the simulated data we know the true reionization history for which the data is produced, we can study with more clarity the possible bias in parameter estimation that is introduced by incorrect assumption about the reionization history. The data is simulated in three channels with frequencies (100 GHz, 143 GHz, 217 GHz) and noise levels per Gaussian beam $(\sigma^{T}_{\rm pix})^2 =$ ($46.25$ $\mu$K$^2$, $36$ $\mu$K$^2$, $17.6$ $\mu$K$^2$) (with $\sigma_{\rm pix}^P =\sqrt{2}\sigma_{\rm pix}^T$ for polarization spectra).  The Full Width at Half Maximum (FWHM) of the three channels are $\theta_{\rm fwhm} =$ (9.5',7.1', 5.0') and we assume a sky coverage fraction of 0.65 \cite{Planck:2006aa}. 

\section{Results}
\label{sec:results}
We present the results of our MCMC analysis as 2-dimensional contour plots on base cosmological parameters. 
\begin{figure*}
\centering
$\begin{array}{ccc}
\subfigure[]{
\includegraphics[width=0.32 \textwidth,clip]{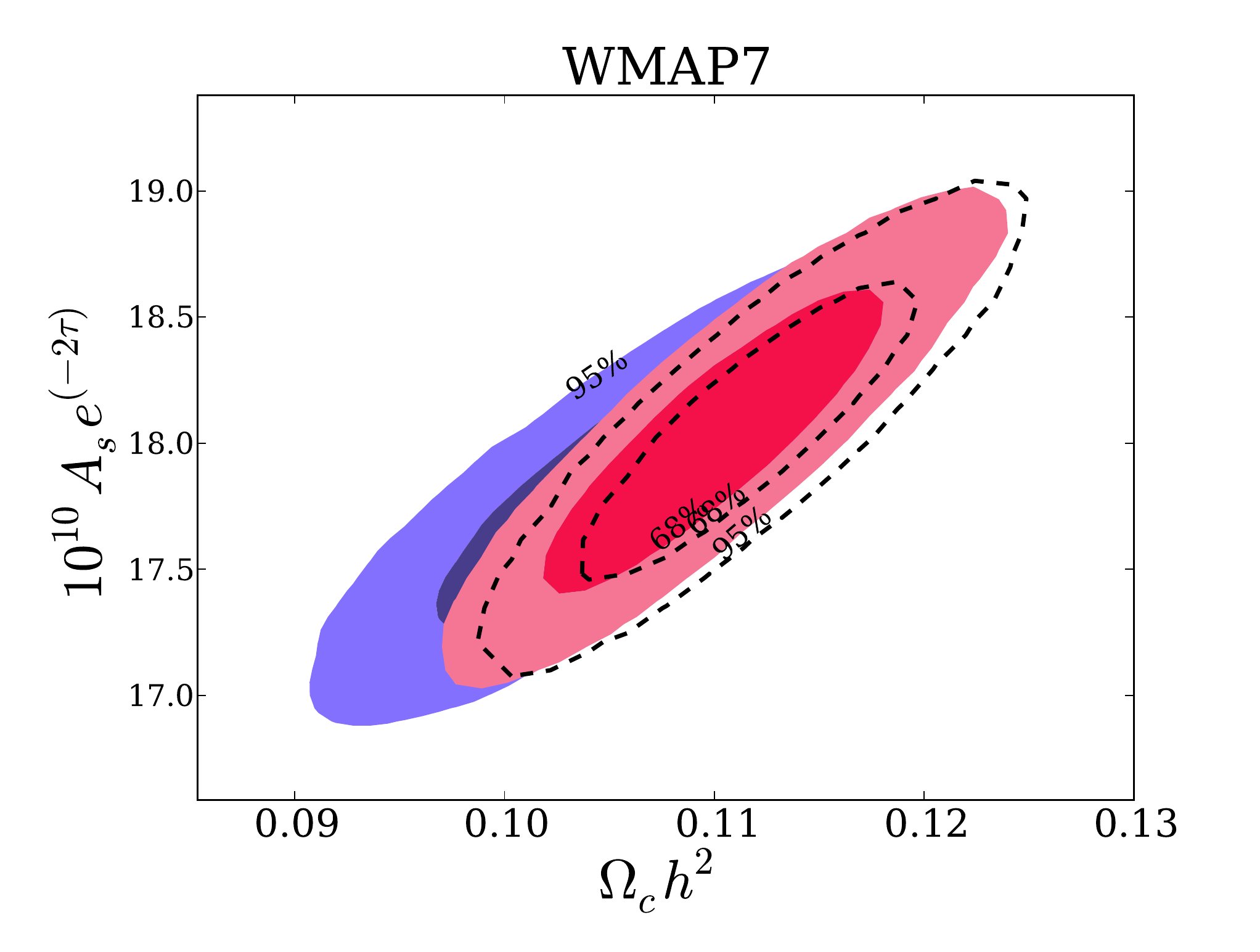}}
\subfigure[]{
\includegraphics[width=0.32 \textwidth,clip]{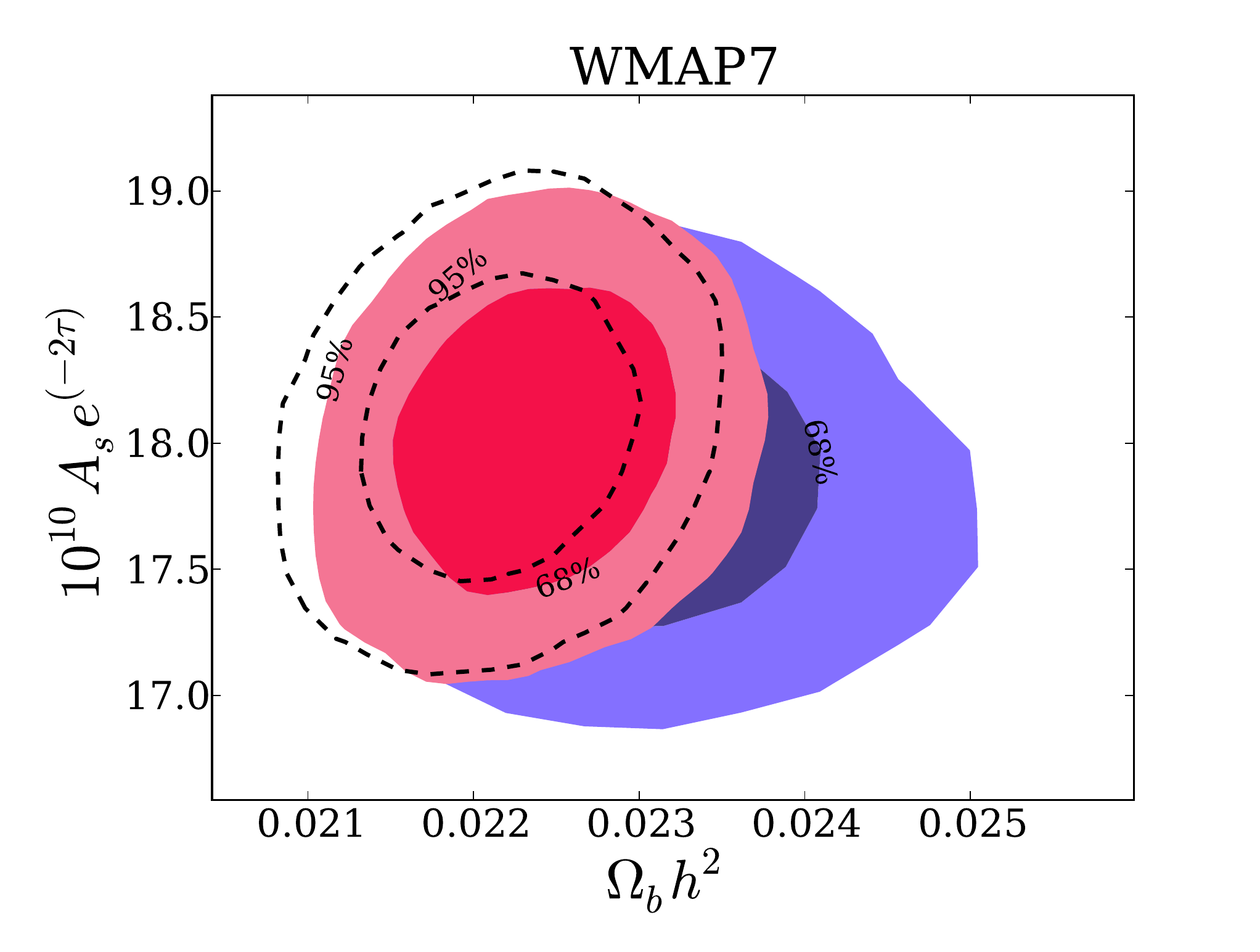}}
\subfigure[]{
\includegraphics[width=0.32 \textwidth,clip]{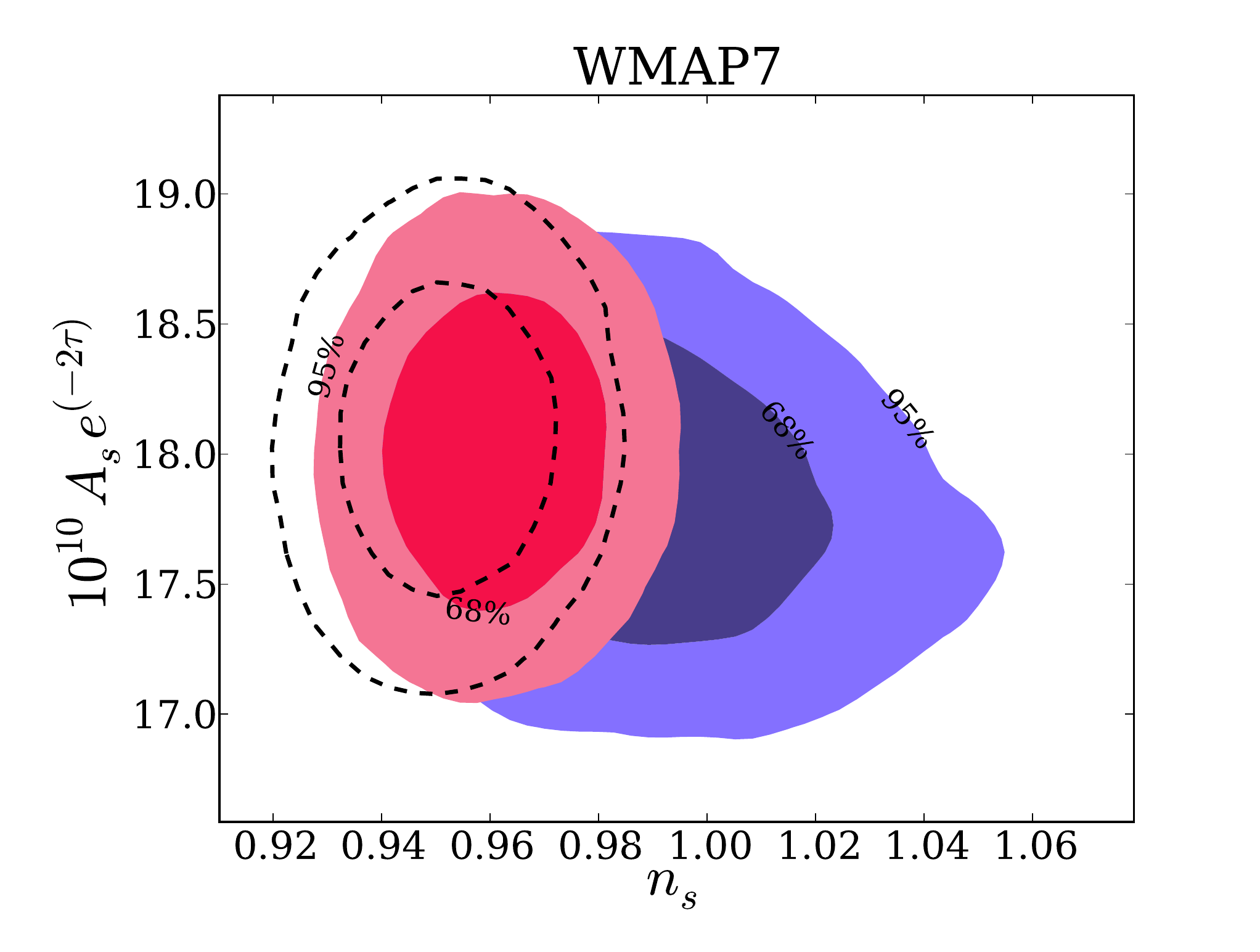}}
\end{array}$
\newline
$\begin{array}{ccc}
\subfigure[]{
\includegraphics[width=0.32 \textwidth,clip]{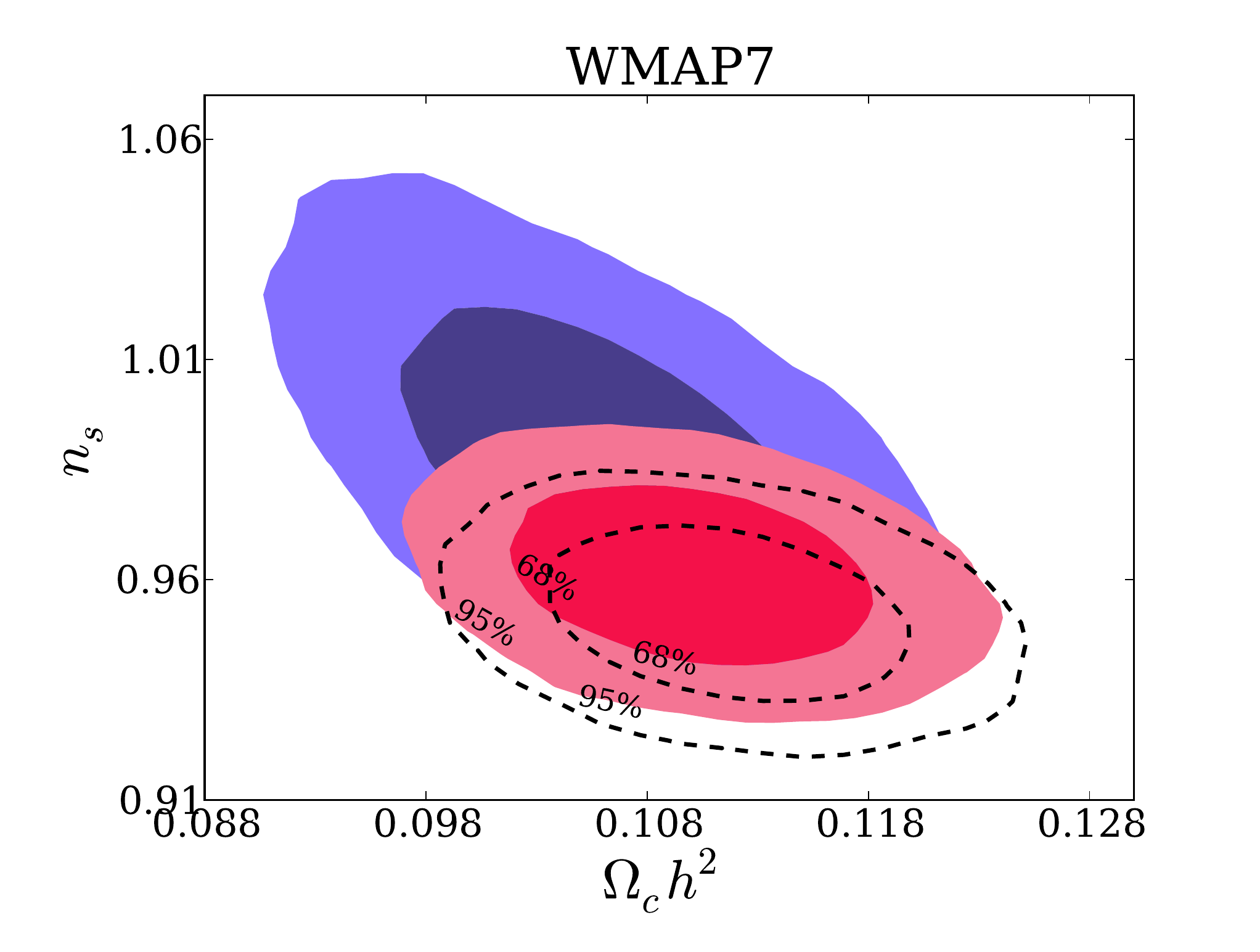}}
\subfigure[]{
\includegraphics[width=0.32 \textwidth,clip]{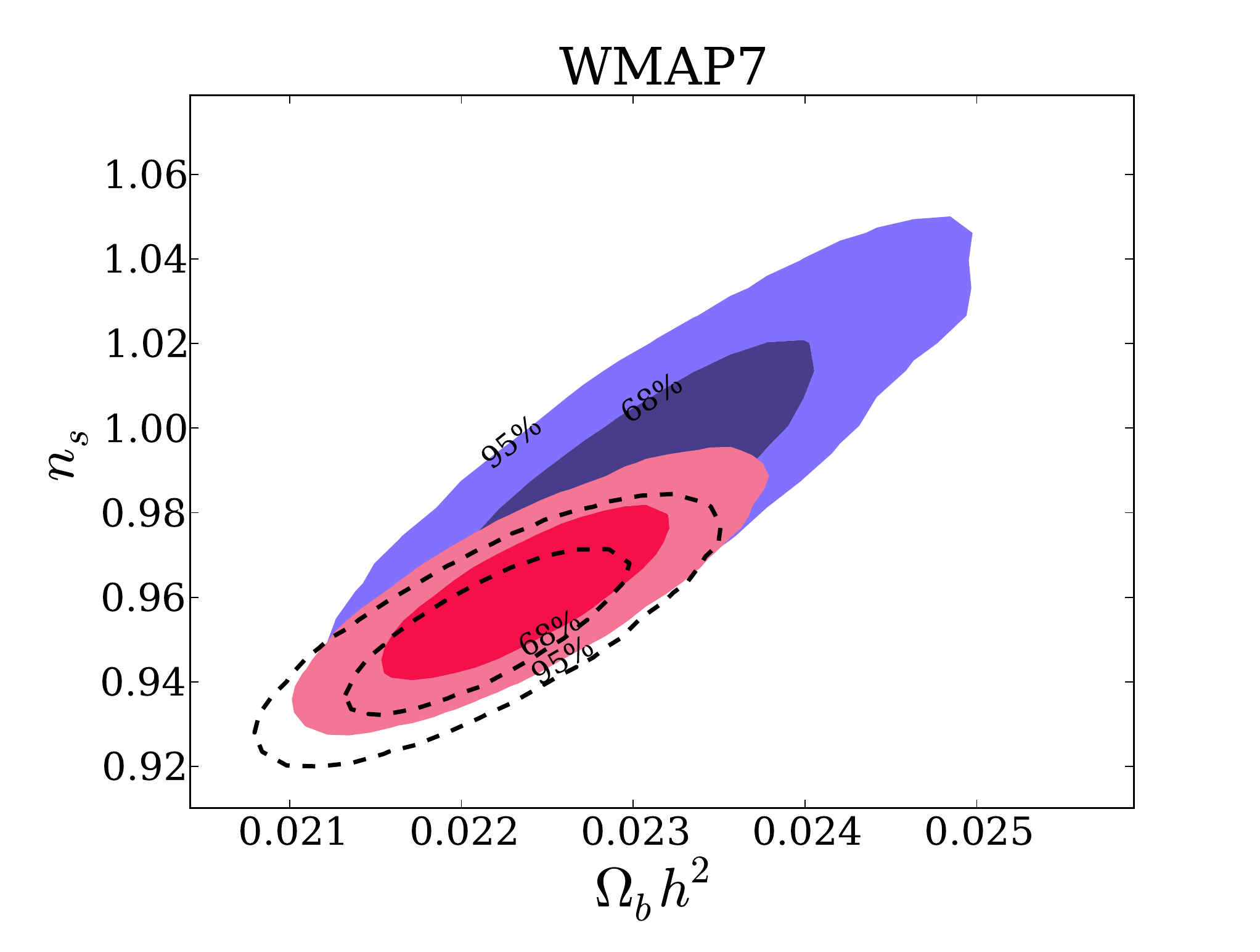}}
\subfigure[]{
\includegraphics[width=0.32 \textwidth,clip]{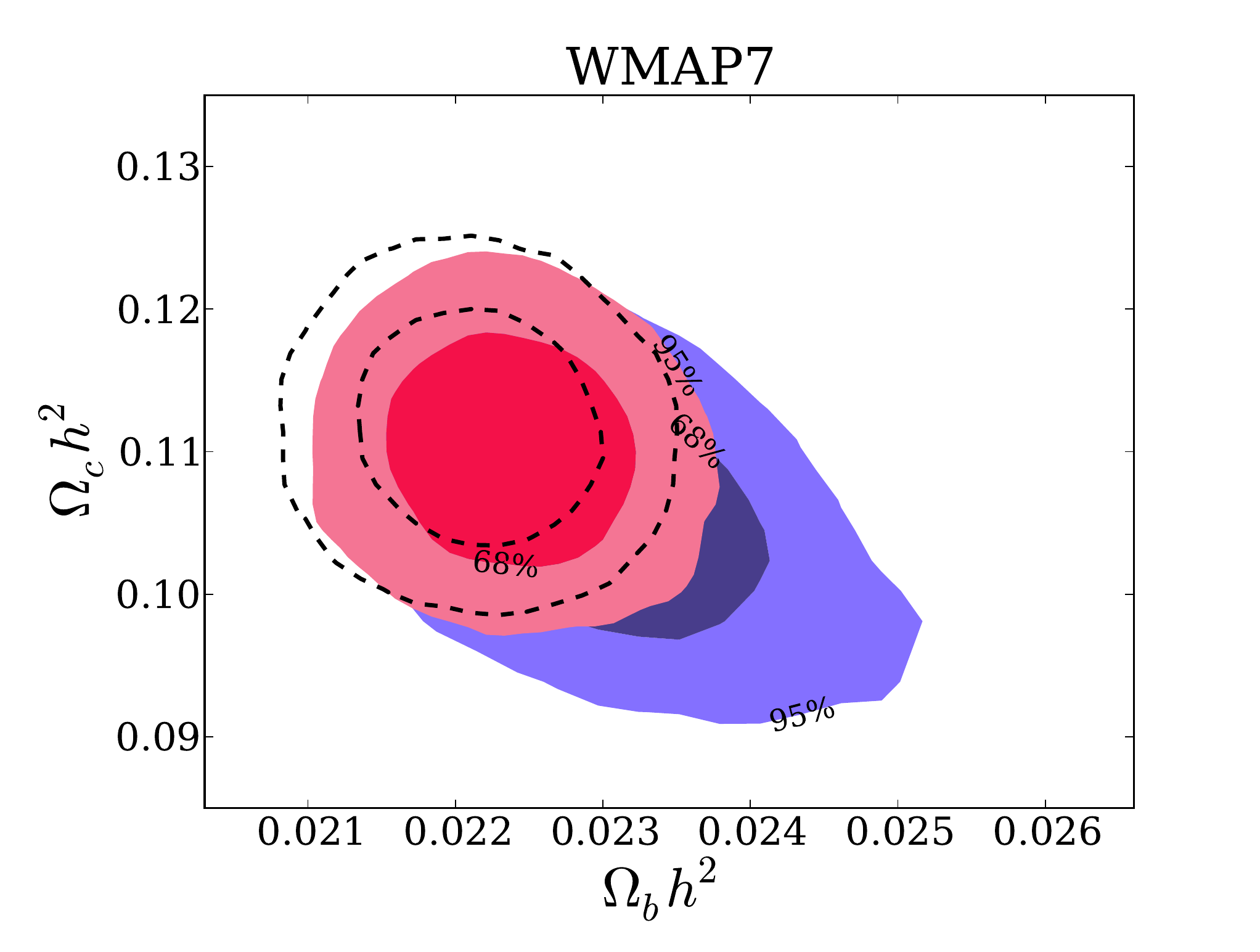}}
\end{array}$
\caption{Marginalized constraints on cosmological parameters for WMAP7 data set. Shaded blue contours correspond to PC reionization while shaded red contours correspond to sudden reionization and dotted contours are obtained assuming the VG reionization history.}     
\label{fig:WMAP7}
\end{figure*} 
\begin{figure*}
\centering
$\begin{array}{ccc}
\subfigure[]{
\includegraphics[width=0.32 \textwidth,clip]{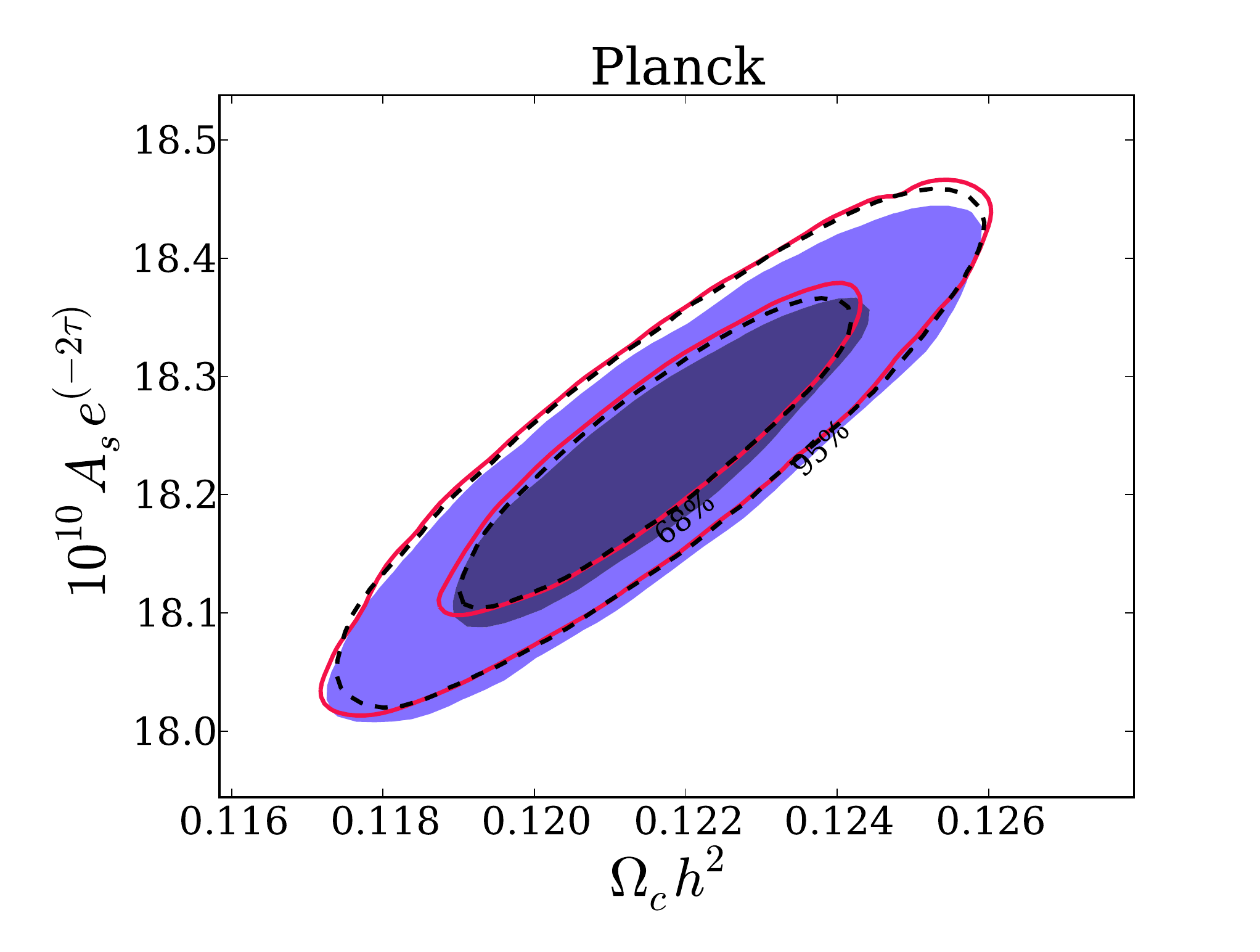}}
\subfigure[]{
\includegraphics[width=0.32 \textwidth,clip]{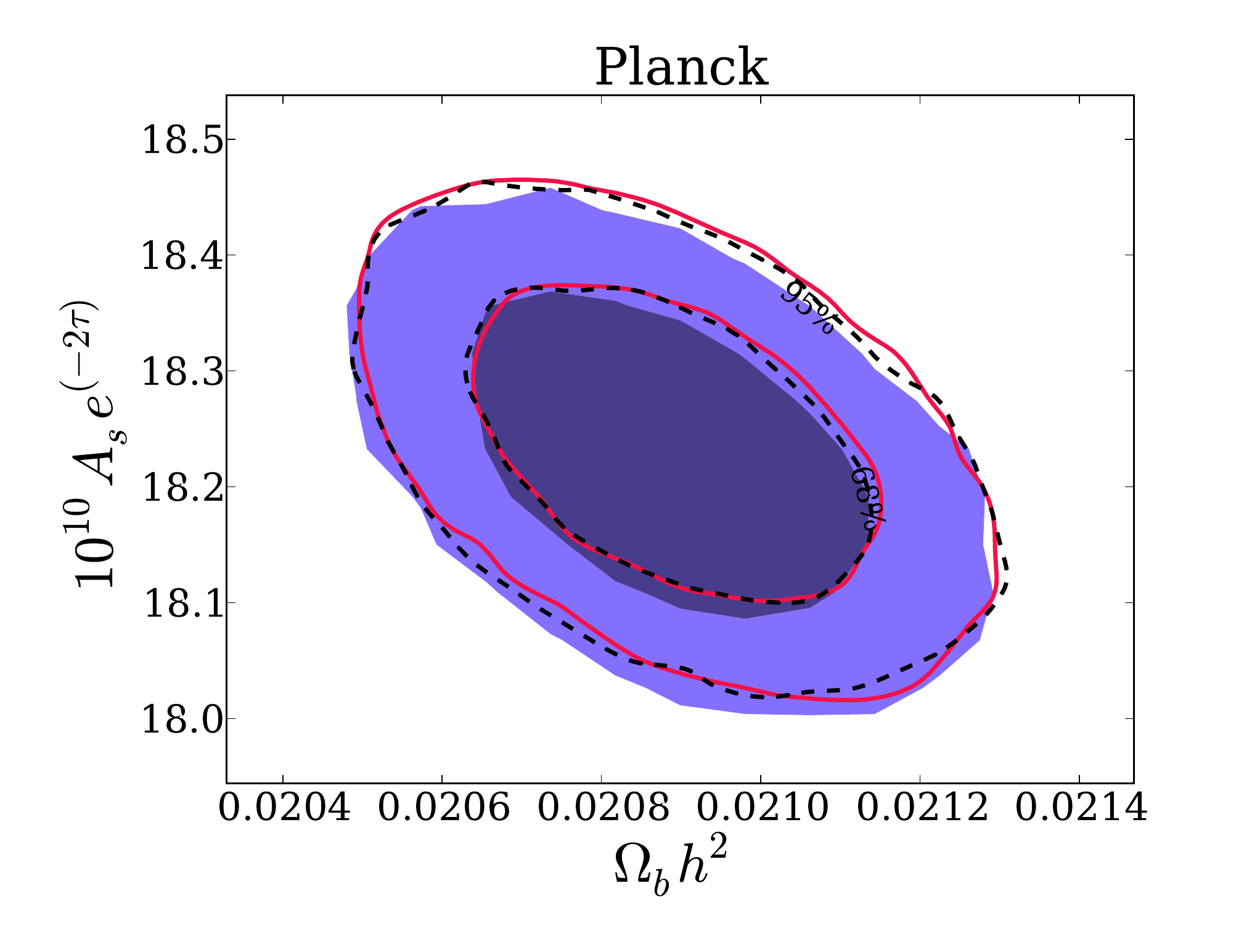}}
\subfigure[]{
\includegraphics[width=0.32 \textwidth,clip]{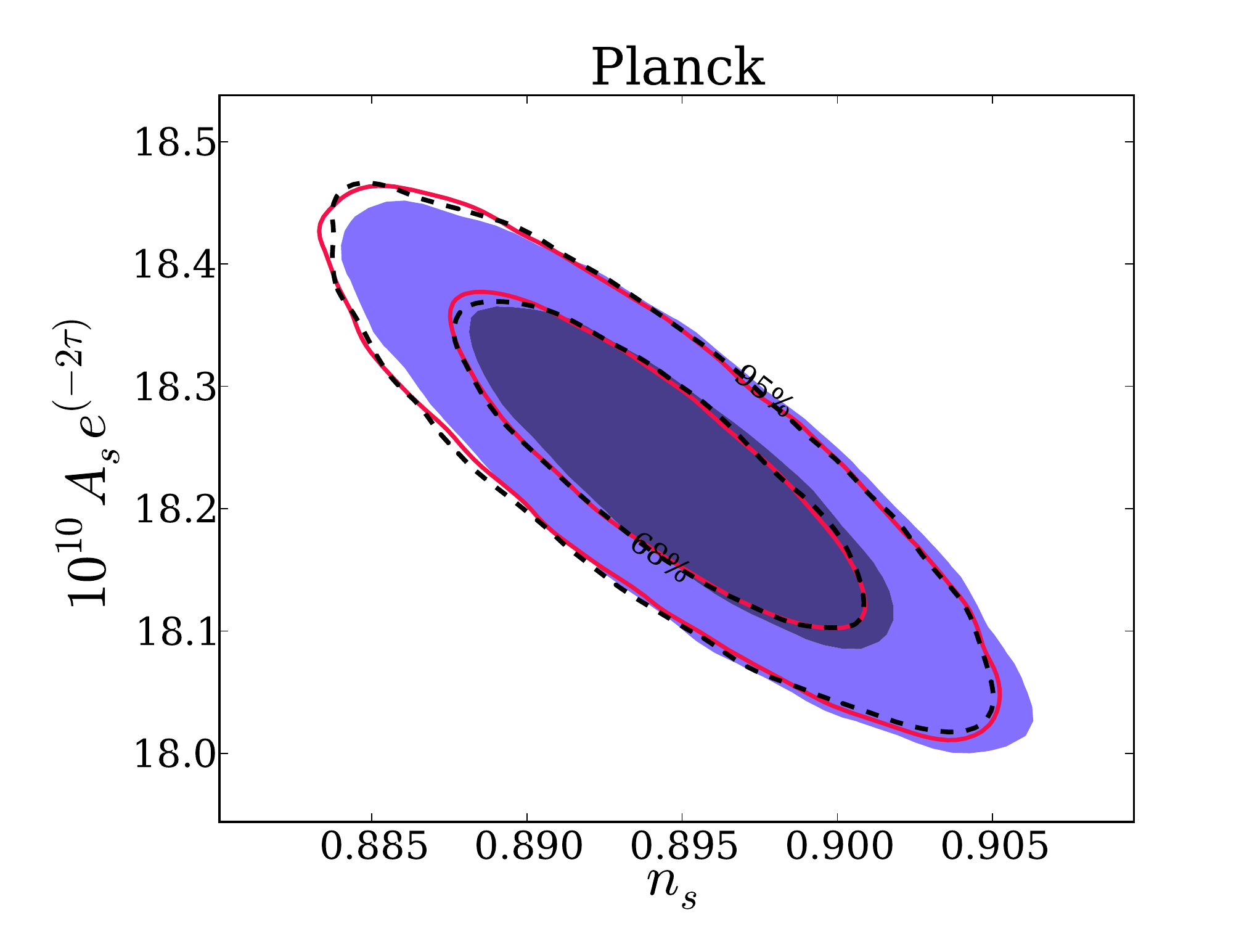}}
\end{array}$
$\begin{array}{ccc}
\subfigure[]{
\includegraphics[width=0.32 \textwidth,clip]{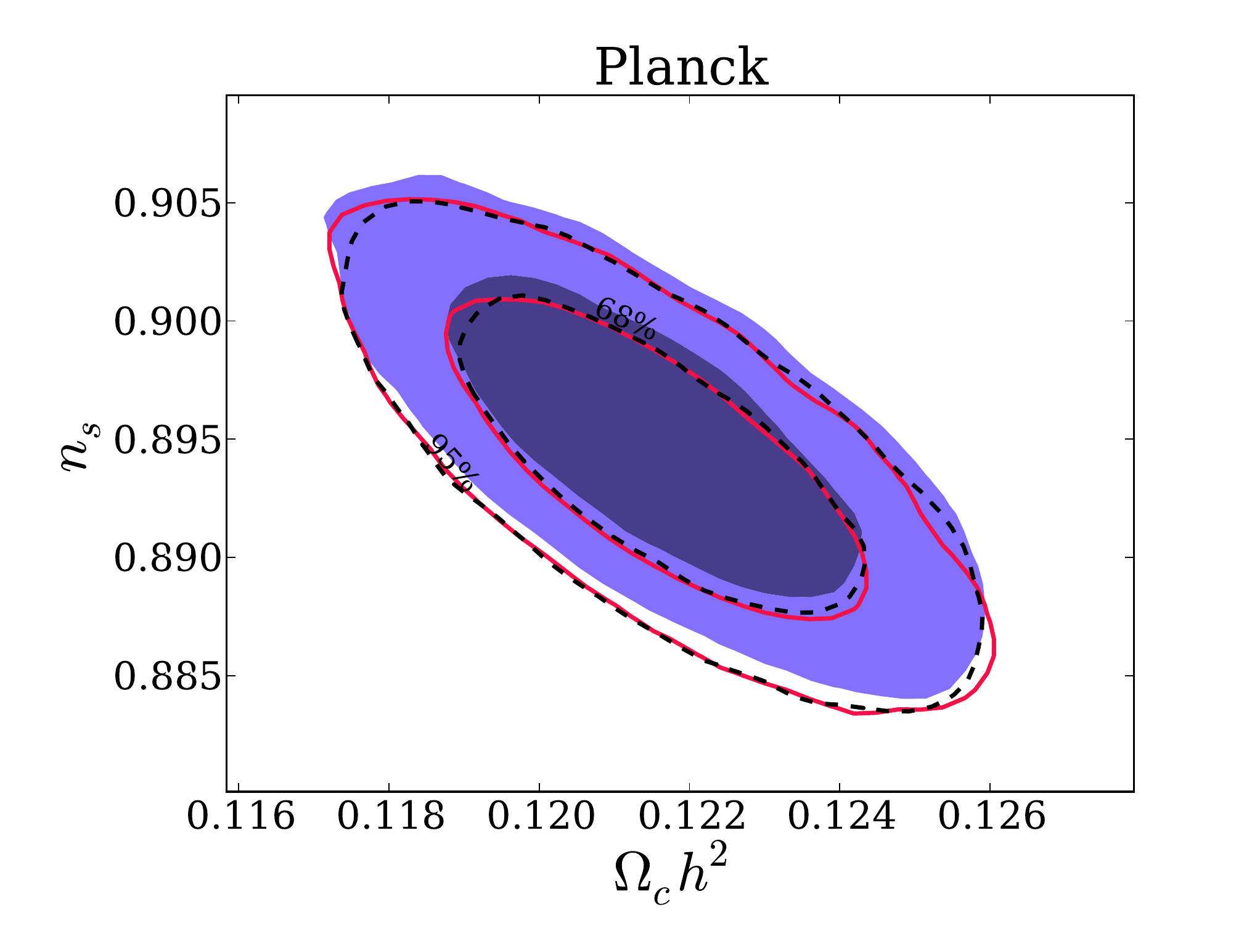}}
\subfigure[]{
\includegraphics[width=0.32 \textwidth,clip]{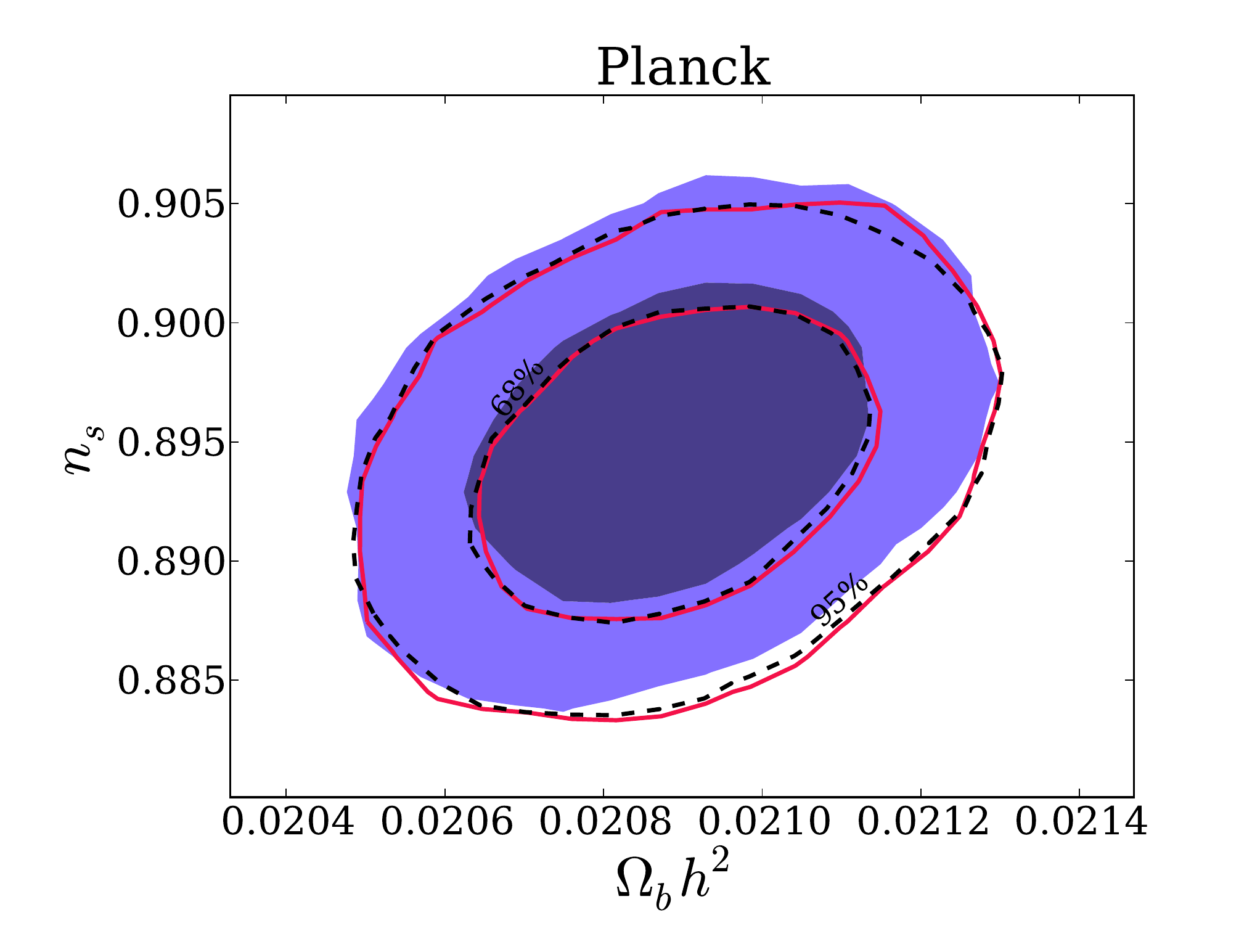}}
\subfigure[]{
\includegraphics[width=0.32 \textwidth,clip]{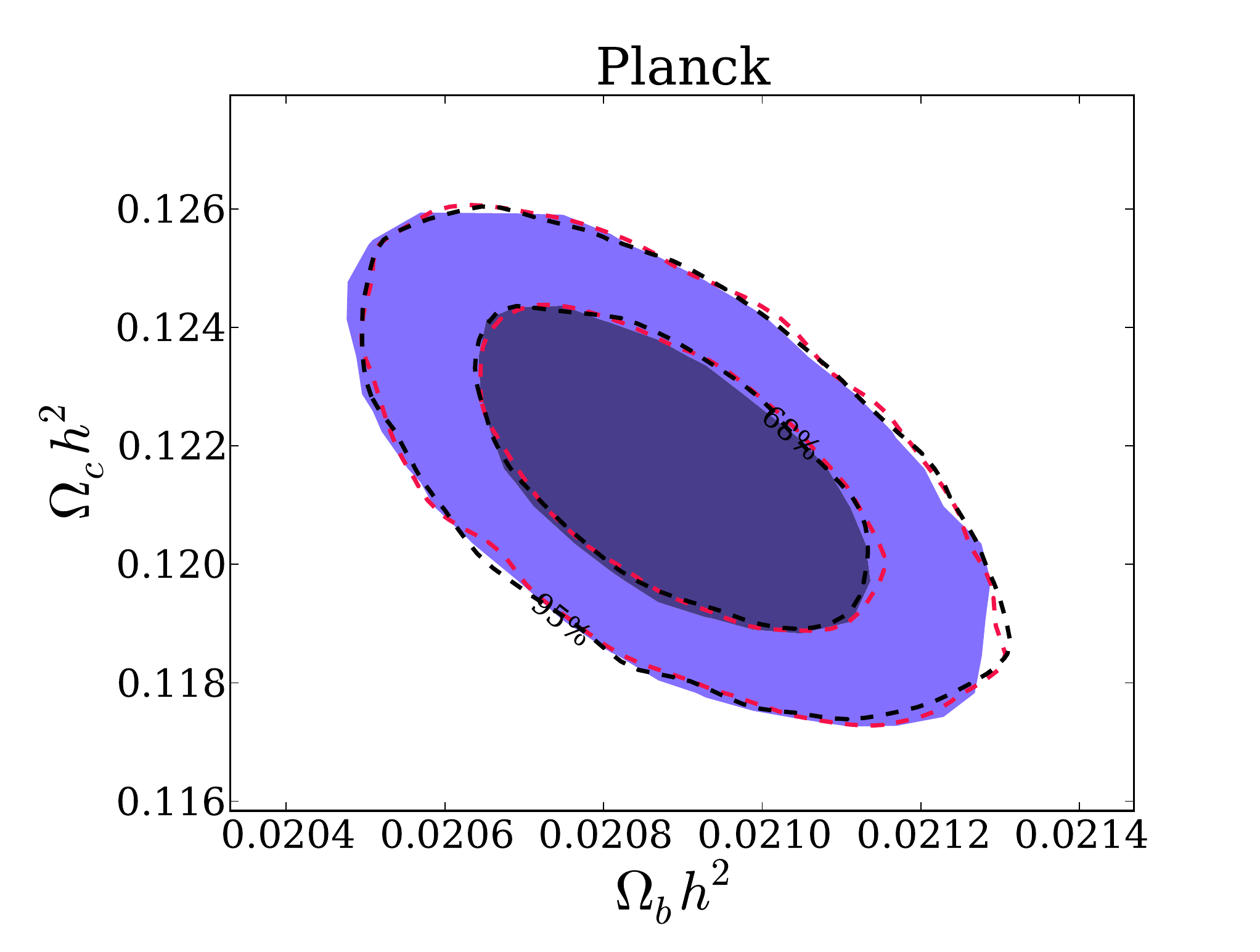}}
\end{array}$
\caption{Marginalized constraints on cosmological parameters for Planck precision CMB data set. Shaded blue contours correspond to PC reionization while solid red contours correspond to sudden reionization and dotted contours are obtained assuming the VG reionization history.}     
\label{fig:Planck}
\end{figure*} 
The constraints on derived parameter $\sigma_8$ and on cosmological parameters for WMAP7 data are shown in Figs. \ref{fig:sigma8-ns} (a) and \ref{fig:WMAP7} respectively. The corresponding constraints for simulated Planck data are shown in Figs. \ref{fig:sigma8-ns} (b) and \ref{fig:Planck}. The dotted black lines correspond to VG reionization, while the filled blue contours are obtained assuming the PC reionization. In WMAP7  plots, the filled red contours correspond to sudden reionization while in Planck plots, the corresponding constraints are shown with solid red contours. 

\begin{figure}
\includegraphics[width=0.5\textwidth,clip]{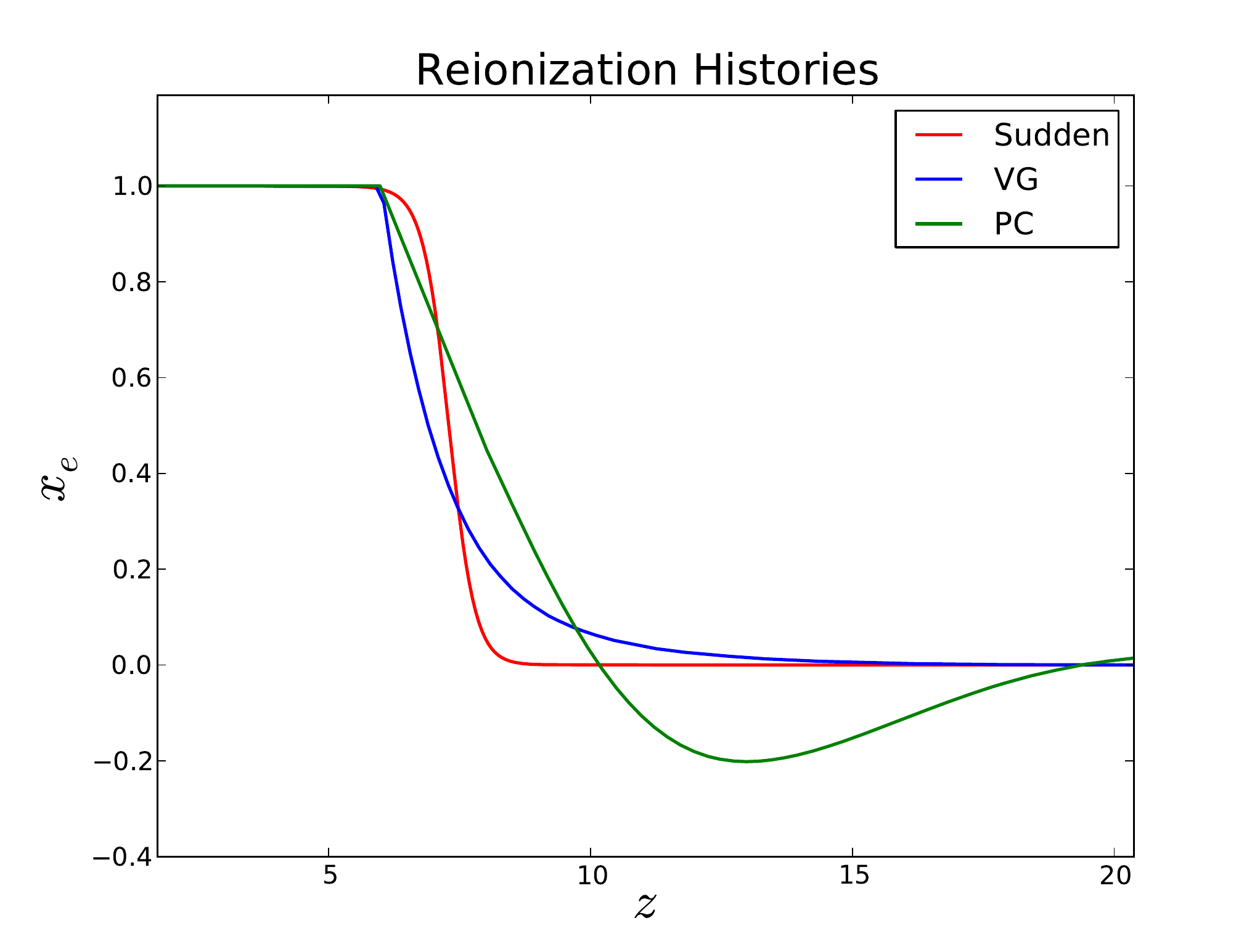}
\caption{Reconstructed reionization history from the first five principal components using simulated  Planck data shown in green while sudden and VG reionization histories are shown in red and blue correspondingly.}
\label{fig:PCA-reion}
\end{figure}

For WMAP7 data, assuming VG reionization history results in a smaller value of $\tau$ compared to the sudden and PC reionizations. Due to the degeneracy between $\tau$ and $A_s$, a lower value of $\tau$ results in smaller values of $A_s$. Our results indicate that for WMAP7, constraints on base cosmological parameters and the derived parameter $\sigma_8$ are mildly affected by the assumption about the reionization history (the shift is less than $1 \sigma$).  Assuming a general reionization history and truncating the expansion at the fifth principal components of $x_e$, degrades the constraints on cosmological parameters. Nevertheless the constraints are consistent with those obtained assuming sudden reionization or the VG reionization. 

One would expect that the enhanced precision of Planck data would distinguish between different reionization models since they would have an impact on the constraints on cosmological parameters. However our results indicate that for reionization histories that are not very different than instantaneous reionization, even at the sensitivity level of upcoming Planck data, parameter estimation from the CMB is only affected by the value of optical depth and not the details of the reionization history. Figures  \ref{fig:sigma8-ns} (b) and \ref{fig:Planck} show the constraints on cosmological parameters for three reionization histories from simulated Planck-quality data. The black dotted contours show the constraints assuming the true reionization history (VG reionization) for which the data is simulated. These results indicate that even at the level of Planck data, constraints on cosmological parameters are not affected by the assumed reionization history. Comparison between the filled blue contours and solid red contours indicates that even for Planck data, the constraints obtained assuming sudden reionization are as good as those obtained assuming general reionization. 
 
Moreover, our results indicate that principal component analysis does not allow for accurate reconstruction of reionization history of the universe from CMB data.  Fig. \ref{fig:PCA-reion} shows the reconstructed reionization history from Planck data using PCs in addition to the corresponding sudden and VG reionization. The reconstructed reionization using PCs exhibits oscillatory behavior at $z>10$ while the true reionization (VG) is a a monotonically decreasing function of redshift. 

\section{conclusions}
\label{sec:conclusions}
The epoch of reionization is the last major phase transition in the history of the universe, which sets a screen between us and CMB. Due to the complexity of the reionization process and lack of direct observational data, in analyzing CMB data, it is commonly assumed that reionization is a sharp transition occurring at a given redshift. Given the enhanced precision of upcoming Planck data, understanding the impact of this unphysical and simplistic assumption is of great relevance.   

In this paper we investigated the question of whether the assumption of sudden reionization introduces any bias into constraints on other cosmological parameters. We conducted three sets of MCMC analysis assuming three different reionization histories:  sudden reionization, a physically motivated reionization history (Sec. \ref{sec:reionmodel}), and a general reionization parameterized by principal components (Hu $\&$ Holder, Mortonson $\&$ Hu)  we compared the constraint on cosmological parameters obtained for these three reionization histories. Our results can be summarized as follows: 
\begin{itemize}
\item For reionization histories which are not drastically different from instantaneous reionization, such as the physically motivated model from Gnedin, {\it{et al.}} simulations (Fig.\ref{fig:Nick-reion}), constraints on cosmological parameters are not sensitive to details of reionization history (Figs. \ref{fig:WMAP7} - \ref{fig:Planck}). Therefore lack of precise knowledge of true reionization history and assuming sudden reionization model does not bias parameter estimation from CMB. 
\item Principal component parametrization of reionization history does not allow for accurate reconstruction of  ionized fraction. 
\end{itemize}
This result is fully consistent with previous results in the literature, in particular Columbo and Pierpaoli \cite{Colombo:2008jr}, who performed a similar investigation using a fiducial two-step reionization model instead of the more physically motivated fiducial model used here. Their results indicate that if the fiducial model is not significantly different from sudden reionization, assuming sudden reionization does not introduce any bias into parameter estimation. For a general reionization history, this assumption would introduce a bias of 1-3 sigma into values of optical depth, amplitude of scalar perturbations and tensor to scalar ratio.  Since we expect the true reionization history to be a smooth function of redshift, the conclusion is that for near-future CMB data sets, sophisticated modeling and reconstruction of the epoch of reionization is unnecessary for the accurate determination of other cosmological parameters of interest. 

\section*{Acknowledgements}
This research is supported  in part by the National Science Foundation under grant NSF-PHY-1066278. WHK thanks the Kavli Institute for Cosmological Physics, where part of this work was completed, for generous hospitality.

\bibliography{SCINC}

\begin{thebibliography}{18}
\expandafter\ifx\csname natexlab\endcsname\relax\def\natexlab#1{#1}\fi
\expandafter\ifx\csname bibnamefont\endcsname\relax
  \def\bibnamefont#1{#1}\fi
\expandafter\ifx\csname bibfnamefont\endcsname\relax
  \def\bibfnamefont#1{#1}\fi
\expandafter\ifx\csname citenamefont\endcsname\relax
  \def\citenamefont#1{#1}\fi
\expandafter\ifx\csname url\endcsname\relax
  \def\url#1{\texttt{#1}}\fi
\expandafter\ifx\csname urlprefix\endcsname\relax\def\urlprefix{URL }\fi
\providecommand{\bibinfo}[2]{#2}
\providecommand{\eprint}[2][]{\url{#2}}

\bibitem[{\citenamefont{Zahn et~al.}(2005)\citenamefont{Zahn, Zaldarriaga,
  Hernquist, and McQuinn}}]{Zahn:2005fn}
\bibinfo{author}{\bibfnamefont{O.}~\bibnamefont{Zahn}},
  \bibinfo{author}{\bibfnamefont{M.}~\bibnamefont{Zaldarriaga}},
  \bibinfo{author}{\bibfnamefont{L.}~\bibnamefont{Hernquist}},
  \bibnamefont{and} \bibinfo{author}{\bibfnamefont{M.}~\bibnamefont{McQuinn}},
  \bibinfo{journal}{Astrophys.J.} \textbf{\bibinfo{volume}{630}},
  \bibinfo{pages}{657} (\bibinfo{year}{2005}), \eprint{astro-ph/0503166}.

\bibitem[{\citenamefont{Mortonson and
  Hu}(2008{\natexlab{a}})}]{Mortonson:2007tb}
\bibinfo{author}{\bibfnamefont{M.~J.} \bibnamefont{Mortonson}}
  \bibnamefont{and} \bibinfo{author}{\bibfnamefont{W.}~\bibnamefont{Hu}},
  \bibinfo{journal}{Phys.Rev.} \textbf{\bibinfo{volume}{D77}},
  \bibinfo{pages}{043506} (\bibinfo{year}{2008}{\natexlab{a}}),
  \eprint{0710.4162}.

\bibitem[{\citenamefont{Mortonson and
  Hu}(2008{\natexlab{b}})}]{Mortonson:2008rx}
\bibinfo{author}{\bibfnamefont{M.~J.} \bibnamefont{Mortonson}}
  \bibnamefont{and} \bibinfo{author}{\bibfnamefont{W.}~\bibnamefont{Hu}},
  \bibinfo{journal}{Astrophys.J.} \textbf{\bibinfo{volume}{686}},
  \bibinfo{pages}{L53} (\bibinfo{year}{2008}{\natexlab{b}}),
  \eprint{0804.2631}.

\bibitem[{\citenamefont{Colombo and Pierpaoli}(2009)}]{Colombo:2008jr}
\bibinfo{author}{\bibfnamefont{L.~P.} \bibnamefont{Colombo}} \bibnamefont{and}
  \bibinfo{author}{\bibfnamefont{E.}~\bibnamefont{Pierpaoli}},
  \bibinfo{journal}{New Astron.} \textbf{\bibinfo{volume}{14}},
  \bibinfo{pages}{269} (\bibinfo{year}{2009}), \eprint{0804.0278}.

\bibitem[{\citenamefont{Pandolfi
  et~al.}(2010{\natexlab{a}})\citenamefont{Pandolfi, Cooray, Giusarma, Kolb,
  Melchiorri et~al.}}]{Pandolfi:2010dz}
\bibinfo{author}{\bibfnamefont{S.}~\bibnamefont{Pandolfi}},
  \bibinfo{author}{\bibfnamefont{A.}~\bibnamefont{Cooray}},
  \bibinfo{author}{\bibfnamefont{E.}~\bibnamefont{Giusarma}},
  \bibinfo{author}{\bibfnamefont{E.~W.} \bibnamefont{Kolb}},
  \bibinfo{author}{\bibfnamefont{A.}~\bibnamefont{Melchiorri}},
  \bibnamefont{et~al.}, \bibinfo{journal}{Phys.Rev.}
  \textbf{\bibinfo{volume}{D81}}, \bibinfo{pages}{123509}
  (\bibinfo{year}{2010}{\natexlab{a}}), \eprint{1003.4763}.

\bibitem[{\citenamefont{Pandolfi
  et~al.}(2010{\natexlab{b}})\citenamefont{Pandolfi, Giusarma, Kolb, Lattanzi,
  Melchiorri et~al.}}]{Pandolfi:2010mv}
\bibinfo{author}{\bibfnamefont{S.}~\bibnamefont{Pandolfi}},
  \bibinfo{author}{\bibfnamefont{E.}~\bibnamefont{Giusarma}},
  \bibinfo{author}{\bibfnamefont{E.~W.} \bibnamefont{Kolb}},
  \bibinfo{author}{\bibfnamefont{M.}~\bibnamefont{Lattanzi}},
  \bibinfo{author}{\bibfnamefont{A.}~\bibnamefont{Melchiorri}},
  \bibnamefont{et~al.}, \bibinfo{journal}{Phys.Rev.}
  \textbf{\bibinfo{volume}{D82}}, \bibinfo{pages}{123527}
  (\bibinfo{year}{2010}{\natexlab{b}}), \eprint{1009.5433}.

\bibitem[{\citenamefont{Larson et~al.}(2011)\citenamefont{Larson, Dunkley,
  Hinshaw, Komatsu, Nolta et~al.}}]{Larson:2010gs}
\bibinfo{author}{\bibfnamefont{D.}~\bibnamefont{Larson}},
  \bibinfo{author}{\bibfnamefont{J.}~\bibnamefont{Dunkley}},
  \bibinfo{author}{\bibfnamefont{G.}~\bibnamefont{Hinshaw}},
  \bibinfo{author}{\bibfnamefont{E.}~\bibnamefont{Komatsu}},
  \bibinfo{author}{\bibfnamefont{M.}~\bibnamefont{Nolta}},
  \bibnamefont{et~al.}, \bibinfo{journal}{Astrophys.J.Suppl.}
  \textbf{\bibinfo{volume}{192}}, \bibinfo{pages}{16} (\bibinfo{year}{2011}),
  \eprint{1001.4635}.

\bibitem[{\citenamefont{Volonteri and Gnedin}(2009)}]{Volonteri:2009ck}
\bibinfo{author}{\bibfnamefont{M.}~\bibnamefont{Volonteri}} \bibnamefont{and}
  \bibinfo{author}{\bibfnamefont{N.}~\bibnamefont{Gnedin}},
  \bibinfo{journal}{Astrophys.J.} \textbf{\bibinfo{volume}{703}},
  \bibinfo{pages}{2113} (\bibinfo{year}{2009}), \eprint{0905.0144}.

\bibitem[{\citenamefont{Belikov and Hooper}(2009)}]{Belikov:2009qx}
\bibinfo{author}{\bibfnamefont{A.~V.} \bibnamefont{Belikov}} \bibnamefont{and}
  \bibinfo{author}{\bibfnamefont{D.}~\bibnamefont{Hooper}},
  \bibinfo{journal}{Phys.Rev.} \textbf{\bibinfo{volume}{D80}},
  \bibinfo{pages}{035007} (\bibinfo{year}{2009}), \eprint{0904.1210}.

\bibitem[{\citenamefont{Bouwens et~al.}(2007)\citenamefont{Bouwens,
  Illingworth, Franx, and Ford}}]{Bouwens:2007hn}
\bibinfo{author}{\bibfnamefont{R.~J.} \bibnamefont{Bouwens}},
  \bibinfo{author}{\bibfnamefont{G.~D.} \bibnamefont{Illingworth}},
  \bibinfo{author}{\bibfnamefont{M.}~\bibnamefont{Franx}}, \bibnamefont{and}
  \bibinfo{author}{\bibfnamefont{H.}~\bibnamefont{Ford}}
  (\bibinfo{year}{2007}), \eprint{0707.2080}.

\bibitem[{\citenamefont{Bouwens et~al.}(2008)\citenamefont{Bouwens,
  Illingworth, Franx, and Ford}}]{Bouwens:2008hm}
\bibinfo{author}{\bibfnamefont{R.~J.} \bibnamefont{Bouwens}},
  \bibinfo{author}{\bibfnamefont{G.~D.} \bibnamefont{Illingworth}},
  \bibinfo{author}{\bibfnamefont{M.}~\bibnamefont{Franx}}, \bibnamefont{and}
  \bibinfo{author}{\bibfnamefont{H.}~\bibnamefont{Ford}}
  (\bibinfo{year}{2008}), \eprint{0803.0548}.

\bibitem[{\citenamefont{Gnedin et~al.}(2008)\citenamefont{Gnedin, Kravtsov, and
  Chen}}]{Gnedin:2007pw}
\bibinfo{author}{\bibfnamefont{N.~Y.} \bibnamefont{Gnedin}},
  \bibinfo{author}{\bibfnamefont{A.~V.} \bibnamefont{Kravtsov}},
  \bibnamefont{and} \bibinfo{author}{\bibfnamefont{H.-W.} \bibnamefont{Chen}},
  \bibinfo{journal}{Astrophys.J.} \textbf{\bibinfo{volume}{672}},
  \bibinfo{pages}{765} (\bibinfo{year}{2008}), \eprint{0707.0879}.

\bibitem[{\citenamefont{{Shull} and {van
  Steenberg}}(1985)}]{1985ApJ...298..268S}
\bibinfo{author}{\bibfnamefont{J.~M.} \bibnamefont{{Shull}}} \bibnamefont{and}
  \bibinfo{author}{\bibfnamefont{M.~E.} \bibnamefont{{van Steenberg}}},
  \bibinfo{journal}{\apj} \textbf{\bibinfo{volume}{298}}, \bibinfo{pages}{268}
  (\bibinfo{year}{1985}).

\bibitem[{\citenamefont{Ricotti et~al.}(2001)\citenamefont{Ricotti, Gnedin, and
  Shull}}]{Ricotti:2001zg}
\bibinfo{author}{\bibfnamefont{M.}~\bibnamefont{Ricotti}},
  \bibinfo{author}{\bibfnamefont{N.~Y.} \bibnamefont{Gnedin}},
  \bibnamefont{and} \bibinfo{author}{\bibfnamefont{J.~M.} \bibnamefont{Shull}}
  (\bibinfo{year}{2001}), \eprint{astro-ph/0110432}.

\bibitem[{\citenamefont{Lewis and Bridle}(2002)}]{Lewis:2002ah}
\bibinfo{author}{\bibfnamefont{A.}~\bibnamefont{Lewis}} \bibnamefont{and}
  \bibinfo{author}{\bibfnamefont{S.}~\bibnamefont{Bridle}},
  \bibinfo{journal}{Phys.Rev.} \textbf{\bibinfo{volume}{D66}},
  \bibinfo{pages}{103511} (\bibinfo{year}{2002}), \eprint{astro-ph/0205436}.

\bibitem[{\citenamefont{Mortonson and
  Hu}(2008{\natexlab{c}})}]{Mortonson:2007hq}
\bibinfo{author}{\bibfnamefont{M.~J.} \bibnamefont{Mortonson}}
  \bibnamefont{and} \bibinfo{author}{\bibfnamefont{W.}~\bibnamefont{Hu}},
  \bibinfo{journal}{Astrophys.J.} \textbf{\bibinfo{volume}{672}},
  \bibinfo{pages}{737} (\bibinfo{year}{2008}{\natexlab{c}}),
  \eprint{0705.1132}.

\bibitem[{\citenamefont{Komatsu et~al.}(2011)}]{Komatsu:2010fb}
\bibinfo{author}{\bibfnamefont{E.}~\bibnamefont{Komatsu}} \bibnamefont{et~al.}
  (\bibinfo{collaboration}{WMAP Collaboration}),
  \bibinfo{journal}{Astrophys.J.Suppl.} \textbf{\bibinfo{volume}{192}},
  \bibinfo{pages}{18} (\bibinfo{year}{2011}), \eprint{1001.4538}.

\bibitem[{713764()}]{Planck:2006aa}
713764 (\bibinfo{year}{2006}), \eprint{astro-ph/0604069}.

\end{thebibliography}

\end{document}